\documentstyle[aps]{revtex}

\newcommand{\patisalam}{SU(2)_L\times SU(2)_R \times SU(4)_C}
\newcommand{\leftright}{SU(2)_L\times SU(2)_R \times U(1)_{B-L} 
\times SU(3)_C}
\newcommand{\stmd}{SU(2)_L \times U(1)_Y \times SU(3)_C}
\begin{document}

\title{SUPERSYMMETRIC UNIFICATION IN THE LIGHT OF NEUTRINO MASS\footnote
{Invited Talk presented at XIII DAE Symposium on High Energy
Physics, Chandigarh, December 1998 }}

\author{Charanjit S. Aulakh}

\address{Department of Physics, Panjab University,\\
Chandigarh, INDIA\\E-mail: aulakh@panjabuniv.chd.nic.in}


\maketitle
\begin{abstract}
\tightenlines
 We argue that with the discovery of
neutrino mass effects at Super-Kamiokande there is a clear
logical chain leading from the Standard Model through the MSSM
and  the recently developed Minimal Left Right Supersymmetric
models  with a renormalizable
see-saw mechanism for neutrino mass to Left Right symmetric SUSY GUTS :
in particular, $SO(10)$ and $SU(2)_L \times SU(2)_R\times SU(4)_c$.
The progress in constructing such GUTS explicitly is reviewed and their
testability/falsifiability by proton decay measurements emphasized.  
\end{abstract}
\section{Introduction}
\tightenlines
The recent observation by the Super-Kamiokande detector 
 of ``intra-  terrestrial'' flavour
oscillations of muon neutrinos produced in the upper atmosphere
 is the first unambiguous
experimental evidence of fundamental physics beyond the Standard Model of
Particle Physics. It has injected a most welcome element of
constraint into the fevered and ``inspired'' speculative
discourse that has characterized much of theoretical particle
physics for over a decade and pointed in the direction in which 
the theory must be extended . 

In this talk I shall argue, while adopting a minimalist 
 scientific aesthetic (Occam's razor,
verifiability/falsifiability (Bacon/Popper) and ``logical beauty
of Nature ''(Einstein/Dirac) as the hallmarks of scientific
discourse), that this very first experimental  clue taken together with
earlier compelling theoretical motivations points firmly to
Supersymmetric Left Right symmetric theories with a
renormalizable see-saw mechanism for neutrino mass as the likely
candidates for the next level of unification of force and matter.

The very first point to emphasize concerning the lessons of
Super-Kamikande\cite{superk}
 is that it has underlined the ancient wisdom expressed
inimitably in the {\it{Pratyabhinjyahridayam}}\cite{pratya}
 of Kshemraja (Kashmir , {\it{circa}} 950 AC) :

\centerline{{\it{tannana anurupgraahayagraahakbhedaat}}}

which translates as  : `` the universe becomes manifold
by the differentiation of reciprocally adapted subjects and
objects''. That is to say, like any physical theory, the
Standard Model should not be regarded merely ``mathematically''
as a  formally consistent (``axiomatic'') construct which 
incidentally codes all available data. 
Rather the SM should be regarded as an effective field
theory whose renormalizabilty and ``unreasonable accuracy'' are
signals that the current level of limitation ( $E\leq .2 TeV $) on our
experimental probes is far smaller than the scale of new physics
$\Lambda_N$ ; whose effects are therefore suppressed by this small
ratio.

 Recall that in the Standard Model the neutrino is an
``odd-ball'' in the sense that besides being the only
electrically neutral fermion it comes without a partner of
opposite chirality. Therefore the lowest dimensional operator
\cite{Weinabs92} that gives it a mass while respecting the
symmetries of the SM is, schematically,

\begin{equation}
(H^{\dagger} L)^2/M
\label{numasop}
\end{equation}

where H is the scalar Higgs field , L the left handed lepton
doublet and M an unknown mass scale characterizing this
hitherto unknown phenomenon. When $H$ develops a vev the
neutrini acquire Majorana masses $\sim <H>^2/M$. In the absence
of new particles this description of neutrino mass is quite
general. Applying Occam's
razor we shall adopt this minimal and 
general prescription for the incorporation 
of neutrino mass as the necessary extension of the SM
indicated by the evidence for neutrino mass. From the
favoured interpretation \cite{superk}
 of the Super-Kamiokande data as evidence of tau
neutrino mass $\sim 10^{-1.5} eV$ the approximate magnitude of
the scale of the new physics leading to neutrino mass is thus 
$M\sim 10^{14\pm .5}$ GeV . 

The huge ratio $M/M_W$ raises a fundamental difficulty in the
viability of the SM as an effective field theory of fermions,
gauge bosons and a scalar Higgs. As is well known , in  QFT scalar masses
(in
contrast to fermions) receive radiative corrections
which are quadratic in the scale at which the loop
integrations are cut off.  The existence of new physics at the
high scale $M$ to which the scalar Higgs is clearly well coupled 
thus implies that physical masses (such as that of the W and Z
gauge bosons which incorporate the Higgs excitations) $\sim 100
GeV$ receive radiative corrections $\sim M$ and thus require
fine-tuning of bare parameters against these corrections order by
order in loops . This is the crux of the so called
``gauge-hierarchy '' problem. 
While fine-tuning  is perfectly acceptable from a
formal standpoint in the context of QFT,
 it is  profoundly disquieting to our physical
intuition to have to stabilize  the low energy theory against 
corrections from far off scales in this ad-hoc manner. Thus
much effort has been made to devise dynamical symmetry breaking
schemes (Technicolor, top condensation etc) that do not rely on
fundamental scalars . However no satisfactory theory of this
type has been found which can pass the various stringent constraints
imposed by experiment. Therefore the alternative solution
provided by (softly broken) Supersymmetry has gained widespread
acceptance inspite of the fact that, so far, the only ``evidence''
in favour of this scheme is, at best, indirect and based on
unproven hypotheses.

The supersymmetric resolution of the gauge heirarchy problem
invokes the presence of superpartners of opposite statistics but
otherwise mostly identical  quantum numbers for
each of the fields of the standard model. Thus chiral fermions
are accompanied by complex scalars and vice versa while gauge
bosons acquire Majorana fermion partners . Since Bose and Fermi
loop corrections carry opposite signs and the relevant couplings
are related by supersymmetry the troublesome quadratic
dependence on the cut off cancels out leaving a correction of
form 
\begin{equation}
\delta m_{H,W}^2 \sim \alpha_{EW}|m_{Boson}^2 -m_{Fermion}^2| 
\label{masscor}
\end{equation}

where the difference of mass between bosons and fermions is
introduced by terms that break supersymmetry only softly .
Thus softly broken Supersymmetric theories are insensitive to
the details of the physics at much higher scales\cite{kaul} and  enable
one to treat the supersymmetrized SM as a consistent and well defined
effective field theory with local symmetry group $G_{123}$ and a scalar
doublet order parameter : as is consistent with the multitude
of precision tests to which it has been subjected.

Given the indication of new physics at a high scale
$M$ and the need to supersymmetrize the theory in order that it
be structurally stable against disruption by the unknown physics
at some high scale it is natural to ask what might be the  appearance of
the
theory at the large scale $M$. As is well known , in QFT
the parameters of the lagrangian (couplings , masses etc.) should
be taken to be scale dependent so as to control the effects of
large logarithms of ratios of energy scales on the convergence
of the perturbation series. One uses the Renormalization
Group(RG) to run these parameters as a function of the energy
scale($Q$) to obtain the effective coupling at the scale of
interest. Only fields which are light on the scale of
interest contribute to the running upto that scale. Making
the minimal assumption of no other new physics till the scale
$M$,  one can
estimate the changes in the SM couplings (which are now known
fairly accurately : better than 1\% for the Electroweak couplings
and $\sim 5 \%$ for the QCD coupling) . One solves the RG
equations for the gauge couplings  $\alpha_i, i=1,2,3$, 
of the SM gauge group $G_{123}$ : 

\begin{equation}
{d\alpha_i\over {dt}} = {1\over{2 \pi}} b_i \alpha_i^2 
\label{RGeqn}
\end{equation}

here $t= ln Q$ ,and we keep only terms to one loop order in 
gauge couplings and ignore the
(small) effects of the Yukawa couplings of the matter fields which
enter only at two loop order :
For the standard model the one loop coefficients \cite{gqw75}
are ($N_g$ is the number of generations and $N_H$ the number of
Higgs doublets)
\begin{equation}
{\vec{b}}= (0,{-22\over 3},-11) + {4\over 3} N_g(1,1,1) +
N_H({1\over 10},{1\over 6},0)
\label{RGcf1}
\end{equation}

At the time when the calculation of the running was first done
(1975) till the mid 80's the EW couplings were known very
approximately ($Sin^2\theta_W= .215 \pm .014$ as of 1982
 and the top quark mass was thought to be possibly
as low as $20 GeV$ ). The running showed that the three gauge
couplings became equal to within the accuracy permitted by the
low energy data for $N_g=3$ and $N_H=1$ at a scale $Q=M_{GUT}\sim
10^{14.5} GeV$. Notice the coincidence of the unification scale
with what we now strongly suspect to be the scale of the physics
giving rise to neutrino mass. Much will be made of this in what
follows. Furthermore, in 1982 Einhorn
 and Jones \cite{einjn1} and
Marciano and Senjanovic carried out a detailed two loop analysis
of the running for both supersymmetric and non supersymmetric
theories. They found that  two loop
corrections did not substantially alter the agreement with the then
current value of $Sin^2\theta_W$ (read together with the then
prevalent lower limit on the top quark mass $m_t\geq 20 GeV$).
However, the running of the couplings of the supersymmetric version of 
the SM clearly showed that the  slower decrease of the QCD
coupling due to the presence of additional color particles
led to a unification scale $\sim 1.7 \times 10^{16} GeV$ while for
$N_H=2$ (the minimum value allowed by SUSY) the value of
$Sin^2\theta_W$ was as large as $.233$. This result  was apparently
incompatible with the data available at that time .
With remarkable prescience they noted that the effect of the top
quark mass on the $\rho $ parameter of the standard model 
\begin{equation}
\Delta\rho={{3\alpha_{em}m_t^2}\over {16 \pi Sin^2\theta_W m_W^2}} +..
\label{delrho}
\end{equation}
implied that the value of $Sin^2\theta_W$ would rise by $\sim
.017$ if the top quark mass were $\sim 200 GeV$. Thus they
{\it{predicted}} that in such a case the gauge coupling
unification would occur {\it{only}} for the SUSY case while unification
of couplings in the SM with one doublet would be ruled out.
With the availability of precise data from LEP and the
increasingly better lower limits for the top quark mass 
(culminating in its discovery at $\sim 175 GeV$) the conflict
between coupling unification and the Minimal Supersymmetric
Standard Model (MSSM)  steadily ebbed
while it developed and became sharper for the SM. In 1991 the
analysis was redone \cite{amlflo} keeping the new data and its errors in
view
and the predictions of Einhorn,Jones,Marciano and Senjanovic were
convincingly reaffirmed leading to a resurgence of interest in
Grand Unified Theories. The reason for their eclipse in the
nonce having been that the lower scale of unification implied
that the nucleon decay lifetime $\tau_N \sim (M_{X}/g_U)^4 \sim
10^{29\pm 1} yrs$ due to the exchange of SU(5) GUT gauge bosons was in
conflict with the available experimental limits
\cite{kolar,kamiokande} . These also ruled out unification of SM couplings
using larger $N_H$ since those came with even lower values of
$M_X$. 

Given that the gauge couplings beome equal at $M_X$ it becomes
natural to enquire whether the logic of EW unification via
spontaneous breaking of gauge symmetry can be
extended to further unify the three gauge groups of the SM and
thereby explain peculiar clutch of fermion representations into which
known matter is organized within the SM . The remarkable properties of the
tracelessness of the electric charge operator 
$Q=T_{3L} + Y/2$,  and the cancellation of gauge anomalies 
enjoyed by the chiral fermions of each generation of the SM,
fairly cry out for the neat justification provided by embedding
the SM in a GUT . As early as 1974 \cite{ps,gg,fritzsch} it had
been noted that the groups $ G_{224}=G_{PS}=SU(2)_L \times SU(2)_R
\times SU(4)_c,  SU(5),  SO(10)$ were compellingly selected by the
structure of standard model coupled with Occam's razor, while
other possibilities were either more recondite or , since they
included these groups as subgroups, more baroque. Each  of the
three minimal possibilities has certain special virtues and we
shall conduct our discussion within the fairly general framework
offered by them. 

The first and earliest possibility , suggested by Pati and
Salam \cite{ps}, was the seminal idea of Grand Unification : consider the 
SM gauge generators to be linear combinations of the generators
of a larger gauge group which leave the vacuum of the theory
invariant. In other words the larger group is spontaneously
broken at some large scale by the  vevs of suitable Higgs fields.
These vevs are left invariant by the SM generators so
that the effective theory at lower energies had the symmetry of
the SM till it in turn was broken by the SM Higgs vev. Pati and
Salam noted that by promoting leptonicity to the status of a
fourth colour 

i)  The SM gauge group could be embedded in 
$G_{224}=G_{PS}=SU(2)_L \times SU(2)_R\times SU(4)_c$
with $SU(3)_c\in SU(4)_c$ and ${Y\over 2} = T_{3R} + a \lambda_{15}$

ii) the fermion quantum number assignments of the SM then
followed naturally from the identification of
${\hat{Q}}_L(2,2,4) \oplus {\hat{Q}}^c_L(2,2,4)$
with the standard model fermions
$(Q_L,L_L)\oplus (Q^c_L, L^c_L)$ where the antilepton doublet
$L^c_L=(\nu^c_,l^c)_L$ . Notice  the natural introduction of the
chiral partner($\nu^c_L$) of the neutrino $\nu_L$ . As we shall see its
mating with the neutrino, with the additional feature of
majorana masses for each, in the context of the ``see-saw ''
mechanism \cite{gmann}, resolves her odd-ball and ``single-ular''
status. The absence of $\nu^c_L$ from the observed low energy spectrum 
being attributed to both the fact that it is a SM
singlet and so feels no EW gauge force and to the fact that
being a SM singlet nothing protects it from obtaining a large
mass.

ii) The symmetry breaking chain 

\begin{eqnarray}
G_{PS} =\patisalam & \stackrel{M_{PS} }{\longrightarrow} &
\leftright \nonumber \\
& \stackrel{ M_R }{\longrightarrow} &
\stmd 
\end{eqnarray}

can be realized by using the Higgs representaions 
$A(1,1,15), \delta^c(1,3,10),$ $ \delta(3,1,{\overline{10}}) $ and
$\phi(2,2,1)$ with 
\begin{equation}
<A>\sim M_{PS}\qquad <\Delta^c>\sim <{\overline{\Delta}}^c> \sim M_R \qquad
<\phi>\sim M_W
\end{equation}

Note the intermediate stage of symmetry breaking where $G_{PS}$ is broken 
down to the so called ``Left-Right symmetric '' gauge group
$SU(2)_L\times SU(2)_R \times U(1)_{B-L}$
which will be the focus of much of our discussion in what
follows. Note also that the generator $\lambda_{15} \sim
Diag(1,1,1,-3)$ of $SU(4)_c$ is , in appropriate normalization,
nothing but the quantum number $B-L$ so that the electric
charge in the LR model takes the appealing parity symmetric form
 \cite{rabimarsh}

\begin{equation}
Q= T_{3L} + T_{3R} + {{B-L}\over 2}
\label{QLR}
\end{equation}

By imposing a discrete symmetry (effectively parity ) 
which interchanges the two SU(2)
factors one obtains a model in which the maximal P and C
asymmetry of the SM is traced to the spontaneous breaking of the
LR discrete symmetry . As we shall see this can be done while at
the same time making $\nu^c_L$ heavy so that the neutrino's 
``single-ularity'' is intimately tied up with source of the maximal
parity violation inherent in the structure of the standard
model.

 The next possibility considered was $SU(5)$ \cite{gg} 
which has the appealing
feature that it  is a simple group. The SM gauge group 
is a maximal subgroup of $SU(5)$ and  the SM 
families of fermions fit into a single (anomaly free !) pair of
$SU(5)$ representations : 
${\bar{5}}(d^c_L +L) \oplus 10 (U^c, Q_L,e^c_L)$. The
symmetry breaking from $SU(5)$ to $G_{123}$ can be accomplished
by the adjoint {\bf{24}} of $SU(5)$ while the SM higgs doublet can be
embedded in the fundamental {\bf{5}}. The principal point to note
here is that the quantum numbers of the $\nu^c_L$ dictate that
it is a $SU(5)$ singlet so that it is plausibly heavy at whatever
scale $SU(5)$ is broken.

 The final possibility $Spin(10)$ (i.e $SO(10)$ plus fermions)
 is both inclusive
(since it has $G_{PS}$ and $SU(5)$ as subgroups) and
aesthetically appealing since it is a simple group. Further  
a generation of fermions embeds
neatly into a single irreducible representation of $SO(10)$ :
namely the irreducible spinor {\bf{16}} which decomposes under $G_PS$
as precisely $(2,2,4) \oplus (2,2,{\bar{4}})$.

Various symmetry breaking chains such as 

\begin{eqnarray}
SO(10) & \stackrel{M_X }{\longrightarrow} &
G_{PS} =\patisalam \times D_{LR} \nonumber \\
& \stackrel{M_{PS} }{\longrightarrow} &
\leftright \nonumber \\
& \stackrel{M_R }{\longrightarrow} &
\stmd 
\label{chain1lr}
\end{eqnarray}

or 
\begin{eqnarray}
SO(10) & \stackrel{M_X }{\longrightarrow} &
G_{PS}= \patisalam  \nonumber \\
& \stackrel{M_R }{\longrightarrow} &
SU(2)_L \times U(1)_R \times SU(4)_C  \nonumber \\
& \stackrel{M_{PS} (M_{BL}) }{\longrightarrow} &
\stmd 
\label{chain2lr}
\end{eqnarray}

may be realized by using the Higgs multiplets {\bf{45,54,210 ..}} etc.
While the $\Delta$s of the PatiSalam model
(whose vevs give neutrinos Majorana  masses) embed neatly in the
$126/{\overline{126}}$.

Consider next the issue of neutrino mass. The crucial relevant
facts about the neutrino in the SM are that :

{ a)} It is the only
neutral fermion and can hence have a (Majorana) mass  
$ M_{\nu} \nu_L \nu_L $ (in obvious two component notation) without
breaking electromagnetic gauge invariance . However such a mass
term is incompatible with the SM gauge invariance. A Majorana
mass compatible with the SM can arise only via the $d=5$
operator of eqn(\ref{numasop}) or some even more suppressed operator.

{b)} It is unaccompanied
by its chiral partner $\nu^c_L$ in strong contrast to the other
fermions of the SM (we use the left handed anti-particle field
rather than the right handed particle for uniformity of notation
with the SUSY case). 

{c)} At the renormalizable level the SM enjoys an anomaly
free global $U(1)_{B-L}$ symmetry which is violated by the 
operator of eqn(\ref{numasop}). Thus the magnitude of neutrino
mass is closely connected to the strength of $B-L$ violation .

Till recently all experimental data were compatible with all
neutrinos having exactly zero mass. Observations at the $50 kT$ 
water Cerenkov detector SuperKamiokande in a deep mine in
Japan during 1996-98 have shown that muon neutrinos produced in
the upper atmosphere above the antipodes of Super K are depleted
by the time they reach the detector while those from directly
overhead show no deficit. The data are consistent with
oscillations of the upcoming upcoming muon neutrinos into some
other flavour of neutrino ( $\nu_{\tau}$ is favoured) whose mass
differs from that of the muon neutrino by around $10^{-1.5} eV$
and which mixes with it with mass matrix mixing angle such that 
$Sin^2\theta_M \geq .82$. There are also indications that
electron neutrinos emitted by the sun are depleted 
 due to flavour oscillations, into a neutrino with a
mass squared difference of $\sim 10^{-5} - 10^{-6} eV$ 
 {\it{or}} a mass squared
 difference $\sim 10^{-10} eV$ and a large mixing angle.
 Finally the LSND collaboration
reports that reactor muon neutrinos (from pion decays)
possibly oscillate into $\nu_e$ with a mass difference as large as $1 eV$.
However this is in direct conflict with other reactor neutrino
experiments such as KARMEN which show no such effect . The 
explanation of all three effects using neutrino flavour
oscillations probably requires the invocation of a
fourth (``sterile'') flavour of neutrino which is neutral with
respect to the standard model but very light . An experimental
resolution of this controversy should be available in the next
few years. Till that time it would be premature to assume the
necessity of a sterile light neutrino. 

Barring fine tunings , from mass differences in the range
$10^{-2} -10^{-5} eV$, we expect that the neutrino masses lie in
the same range . For instance one could have $m_{\nu_{\tau}}$ of
the order of the Super-K mass difference while
$m_{\nu_{\mu}}\sim 10^{-3} eV >> m_{\nu_e}$ would account for
the Solar neutrino deficit via the MSW mechanism \cite{msw} given suitable
values of the mixing
angles. In any case it is clear that the presence of the
neutrino masses militates for the presence of $B-L$ violating
physics characterized by a scale $\sim 10^{13} - 10^{16} GeV$
which is precisely in the range expected for the Unification
mass in GUT scenarios.

The next question is naturally as to what role the mass scale
$M$ plays in the new physics. The simplest and most natural
explanation is provided by the ``seesaw mechanism'' of
\cite{gmann}. If , like other SM fermions,
the neutrino had a $SU(2)_L$ neutral chiral partner ($\nu^c_L$), 
it , being 
color and charge neutral as well , would be a singlet with
respect to the SM gauge group and might thus naturally take
advantage of its ability to enjoy a Majorana mass term without
breaking charge and pick up a mass characteristic of the high scale of
$B-L$ breaking namely $M$. Moreover once it was present nothing
would prevent the neutrino from pairing with it in a Dirac mass
term so that while respecting the SM gauge symmetry and
renormalizability one could add the following new terms to the
lagrange density :

\begin{equation}
(M_{\nu^c})_{ij} \nu^c_{Li} \nu^c_{Lj} + h_{ij} H^{\dagger} L_i \nu^c_{Lj}
\label{mterms}
\end{equation}

Here the indices $i,j$ run over the three flavours of
neutrinos and $M_{\nu^c}$ is a matrix with eigenvalues $\sim
M\sim M_{B-L}$. From the point of view of low energy physics at
scales $E< 1 TeV$ the observable physics will be coded in the
(non-renormalizable) effective theory obtained by integrating
out the superheavy neutrino states (and momentum components
above that magnitude). This gives precisely the operator of 
eqn(\ref{numasop})  :

\begin{equation}
-h_{ij} (M_{\nu^c}^{-1})_{jk} h_{kl} H^{\dagger} L_i H^{\dagger} L_l
\label{numasopdet}
\end{equation}

So that the suppression of the left neutrino masses is directly
related to the high masses of their chiral partners. This is
called the Type I see-saw mechanism. In case the left neutrinos
acquire (small!) Majorana masses $M_{\nu_L}$ from some other
source then they will add to the above contribution giving the
so called Type II see saw mechanism \cite{rabigoran}.
 This happens quite inescapably in many of
the natural unified models we shall consider and drastically reduces 
their predictivity. For while the Yukawa couplings
$h_{ij}$ are related by the GUT to the masses of quarks the Type
II additional contributions are not similarly constrained by the
low energy data.
Furthermore, one may extend the logic of the SM - where all masses
arise via SSB - and introduce Higgs fields $\Delta^c$ capable of giving the
$\nu^c$ fields Majorana masses when they acquire vevs $\sim M$.
 In that case , given mild
conditions on the Higgs potential, the $B-L$ symmetry of the
Lagrangian will be restored with $B-L(\nu^c)= +1, B-L(\Delta^c)
=-2$. Thus the pattern of neutrino masses will arise from the
spontaneous violation of $B-L$ in such a way that the low energy
theory has a quasi exact $B-L$ symmetry violated only by the
tiny left neutrino Majorana masses. This may nevertheless have
important consequences for fundamental cosmological quantities
such as the baryon to photon ration since dressing of
non-perturbative $B+L$ violating processes by the $B-L$
violating left neutrino masses could erase any $B-L$ (and
therefore $B$) number created at early times \cite{leptogenisis}. 

\section{The R-parity - LR symmetry Connection}

In the above discussion we have argued that 
 the well verified standard model , together with data on
neutrino oscillations and the contextual constraints of the formalism of
effective QFT coupled with the minimalistic scientific
aesthetic that has served us so well, motivate us to consider
supersymmetric versions of the standard model in which the
neutrino mass arises via a ``seesaw'' mechanism . The technology
of supersymmetrizing gauge theories is by now so well known that
we shall not review it here but refer the reader to the
excellent reviews available in the literature \cite{susyrev}.
In the MSSM each of the SM fermion fields $(Q,L,u^c,d^c,e^c)_L $
acquires a complex scalar partner with identical gauge quantum
numbers while the gauge bosons acquire majorana fermion partners.
The single Higgs doublet of the SM must however  be replaced by
at least a pair of doublets of opposite hypercharge together
with their superpartners (Higgsinos) which are chiral fermions
in order to cancel gauge hypercharge anomalies. Gauge
invariant soft supersymmetry breaking terms (scalar masses
$\phi^{\dagger}\phi$ , trilinear scalar gauge invariants and gaugino
masses) allow sufficient freedom to raise superpartner masses
(thus respecting experimental constraints) and at the same time 
preserving the softening of loop divergences which motivated the
introduction of SUSY. 

However the presence of new scalar fields carrying the quantum
numbers of matter fermions destroys one of the most appealing
features of the SM namely the fact that gauge invariance and
renormalizability ensure (perturbatively) exact $B,L$ invariance
of the lagrangian. Since the quasi exactness of these symmetries
is a fixed fact (indeed a prerequisite of our very existence!) 
this serendipitous corollary is not a minor gain in
understanding.   In the MSSM , on the other hand,  there exist a
number of new couplings which violate these crucial invariances.
They may all be derived from the superpotential (we omit a
possible $LH$ term which may be removed by a redefinition of the
Higgs and Lepton fields , and have suppressed all gauge and
generation indices)

\begin{equation}
W_{{\slash R}_p} =\lambda LLe^c + \lambda' LQd^c + \lambda'' u^c
d^c d^c
\label{wrpv}
\end{equation}

These new terms lead to catastrophic proton decay and a host of
other exotic effects which are all known to be severely
suppressed. For instance the  product $\lambda \lambda'$ is
thought to be constrained to be less than $10^{-25}$ by the
absence of nucleon decay. Such small values of the couplings are
very unnatural so that it is appealing to proceed on the minimal
and natural assumption that these parameters are actually exactly zero
(along with the corresponding scalar trilinear couplings in the
soft susy breaking terms). In that case the Lagrangian reaquires
$B-L$ symmetry albeit as an assumption rather than as an
``accidental'' consequence of the model's symmetry and
renormalizability. 

At this point if we recall that neutrino masses are themselves
signals of $B-L$ breaking, we may legitimately wonder if the
MSSM plus neutrino mass would not reintroduce the terms arising
from eqn(\ref{wrpv})  via radiative effects. However this is not so
since the full power of $B-L$ invariance is not required for the
purpose of forbidding these terms.
 It is sufficient to impose only a discrete $Z_2$
symmetry : the so called R parity under which each ``new''
type of field introduced by supersymmetry flips sign. It is a
remarkable fact \cite{rabirpar} that this symmetry is nothing but

\begin{equation}
 R_p =(-1)^{3(B-L)+ 2S}
\label{rbml}
\end{equation}

where S is the field spin . Furthermore, completely equivalently, Matter
parity : $M_p=(-)^{3(B-L)}$ accomplishes the same task without
any loss of generality since the lagrangian is bilinear in fermi
fields and $(-)^{2S} \in Spin(1,3)$ . Now note that the neutrino mass
operator of eqn(\ref{numasop},\ref{numasopdet}) 
is in fact R parity even and thus
will not lead to catastrophic reintroduction of these terms by
radiative corrections. This ad-hoc introduction of R-parity is, however,
quite unsatisfactory, specially since global discrete symmetries
are thought \cite{globgravviol} to be violated by quantum
gravitational effects. On the other hand the connection between
R parity and $B-L$ hints strongly at a deep connection between
these symmetries. The overarching importance of $B-L$ in the
standard model naturally motivates us to take this possibility
seriously.  It is then a pleasant surprise to realize that
in Left-Right symmetric theories $B-L$ is a gauge
symmetry\cite{rabimarsh} and on supersymmetrizing R parity is
therefore a part of the gauge symmetry. Moreover these models
accommodate the seesaw mechanism for neutrino mass in a very
natural and convincing manner which singles them out as the
prime candidates indicated by the discovery of neutrino mass.

In LR symmetric models \cite{lrmodels}
 the gauge group of the SM is extended to
$SU(2)_L\times SU(2)_R\times U(1)_{B-L}\times SU(3)_c$ while the SM
fermions
along with the chiral partner of the left neutrino group into 
the representations $L(2,1,-1,1)\oplus L^c(1,2,1,1)\oplus
Q(2,1,1/3,3)\oplus Q^c(1,2,-1/3,{\bar{3}})$. The  Higgs sector consists
of triplets $\Delta (3,1,2,1)\oplus \Delta^c (1,3,-2,1)$ along
with a bidoublet $\phi(2,2,0,1)$ which contains a pair of SM
doublets with $Y=\pm 1$. Mild conditions on the couplings are
sufficient to implement a discrete symmetry which interchanges
left  with right chiral fermions $SU(2)_L$ with $SU(2)_R$ ,
$\Delta$ with $\Delta^c$ etc.  The system composed of the
former can break $SU(2)_R\times U(1)_{B-L}$ down to $U(1)_Y$ at
some scale larger than $1 TeV$ resulting in a model which
is effectively the SM with the additional feature of a right
handed neutrino. The seesaw mechanism can be implemented most
naturally in this model since gauge invariance and
renormalizability imply that the Yukawa couplings of the Higgs fields
: 
\begin{equation}
h_{ij}^a L_i\phi_a L^c_j + f_{ij} L_i \Delta L_j + f^c_{ij}
L^c_i \Delta^c L^c_j +h.c
\label{yukawas}
\end{equation}

are precisely such as to give a large Majorana mass to the $\nu^c_L$
fields when the $\Delta^c$ field develops a large vev, while the
vev of the $\phi$ field which is dominantly responsible for EW SSB
gives rise to a Dirac mass term between $\nu_L$ and $\nu^c_L$.
Thus a seesaw mechanism occurs very naturally as a
consequence of the hierarchy between $SU(2)_L\times U(1)_Y$ and 
$SU(2)_R\times U(1)_{B-L}$ breaking scales. Since the scalar potential in
general allows couplings of the form 

\begin{equation}
V= M^2 \Delta^2 + \Delta \phi^2 \Delta^c + ......
\label{type2}
\end{equation}

It follows that once $\phi,\Delta^c$ acquire vevs $\sim M_W,M$
respectively $\Delta$ acquires a vev due to the linear term generated :
\begin{equation}
<\Delta> \sim M_W^2/M
\label{deltvev}
\end{equation}

so that the seesaw is in general of Type II.

Finally it bears mention that doublets $\chi(2,1,1,1) \oplus
\chi^c(1,2,-1,1)$  may be used instead of triplets  to break the
left-right symmetry . The price one must pay is that the
triplets with $B-L=\pm 2$ required to implement the see-saw
mechanism must now be composites of the $\chi,\chi^c$ fields so
that the required terms are of dimension 5 and therefore
non-renormalizable. Such models develop other ugly features once
supersymmetry is introduced. In particular they violate R-parity
maximally thus destroying part of the motivation for studying LR
supersymmetric models. We shall not consider these models
further in this talk and refer the reader to the literature 
where they have been discussed exhaustingly \cite{babuetc}.

\section{Minimal SUSY LR Models}

Recapitulating: the SM needs SUSY and the MSSM needs R/M-parity to
be consistent with experiment . R/M-parity is essentially the
$(-)^{3(B-L)}$ global subroup of $U(1)_{B-L}$ symmetry, which is
the only anomaly free continuous global symmetry of the SM and
remains a symmetry of the MSSM if R-parity is imposed.  Thus
in SUSY theories which gauge B-L -among which the most
natural and appealing are LR symmetric theories - R/M-parity is
automatic at the level of the Lagrangian. Moreover these
theories naturally incorporate the seesaw mechanism which
explains the smallness of left neutrino masses as a consequence of the
decoupling of their heavy (SM singlet) chiral partners. In such
theories, when the seesaw is renormalizable i.e the large
Majorana masses of the $\nu^c_L$ fields come from the vevs of
fields $\Delta^c$ which carry $B-L=-2$ the breaking of $B-L$
symmetry by the $<\Delta^c>$ still preserves R-parity. Thus the
only possible source of R-parity violation in the low energy
theory is a vev for  the scalar partner of the $\nu_L,\nu^c_L$ fields
\cite{am82}.

A significant advance \cite{abs} has been the realization that
when -as is generically the case and as is now experimentally
indicated - the scale of $B-L$ violation is $>>M_W$,
phenomenological constraints and the structure of the SUSY
vacuum ensure that R-parity is preserved. 
Viable Minimal LR Supersymmetric Models (MSLRM) have been
constructed in detail \cite{abs,ams,amrs}
 and embedded in GUTS while retaining these
appealing properties.

  Since the argument for R-parity exactness
is so simple and general we present it first in isolation before
going on to the details of the MSLRMs. Given $M_{B-L} >> M_W\sim
M_S$ (the scale of SUSY breaking) it immedeately follows that
the scalar partners of the $\nu^c_L$ fields also have positive
mass squares $\sim M_{B-L}^2 $ and hence are protected from
getting any vevs modulo effects suppressed by these large
masses. Thus when the $\nu^c_L$ superfield is integrated out the
effective theory is the MSSM with R-parity and with $B-L$
violated only by (the SUSY version of) the highly suppressed
operator of eqn(\ref{numasop}). As a result if , for any reason,
the scalar ${\tilde\nu}_L$ were to obtain a vev \cite{am82} the low energy
theory would contain an almost massless scalar (i.e a  ``doublet''
(pseudo) Majoron) in its spectrum. However since such a
pseudo-Majoron couples to the SM gauge fields like its superpartner the
 neutrino \cite{am82} it 
would make an appreciable contribution to the width of the $Z$ gauge boson 
and  is therefore conclusively
ruled out by the very precise measurements of this quantity at
LEP (for essentially the same reasons that the Z width limits
the number of light neutrinos to be 3). It is worth emphasizing
that these arguments being based on decoupling are very robust.
In particular note that the Majoron/longitudinal B-L mode
expected from the spontaneous violation of a global/local $B-L$
symmetry at the large scale $M_{B-L}$ will always decouple. The
pseudo-majoron referred to would arise because {\it{as a consequence
of the R-parity of the model}} the low energy theory has a quasi
exact B-L symmetry which becomes exact in the limit
$m_{\nu_L}\rightarrow 0$ or equivalently $M_{B-L}\rightarrow
\infty$. In fact it is easy to check that the leading
contribution to the mass of this pseudo-Majoron is $M_J^2 \sim
M_S m_{\nu}$ which is far smaller than $M_Z$. Thus it is not
possible to evade this argument by making the doublet Majoron
heavy as has been suggested in the literature \cite{comelli}.
It is also easy to check that the vev for ${\tilde\nu}^c_L$
induced via trilinear scalars in the soft susy breaking
potential ($M_S <LH>L^c +..$), after both $H,{\tilde\nu}_L$
obtain vevs ,is so small ($<{\tilde\nu}^c_L \sim M_S M_W <{\tilde\nu}_L>/
M_{{\tilde\nu}^c}^2$) that it gives a contribution to the
pseudo-Majoron mass 

\begin{equation}
m_J \sim  \frac{ M_S m_{\nu}}{M_W}
\end{equation}

which is even less effective in evading the Z-width bound. To sum up, quite
generally\cite{amrs}  :

 {\it{The low energy effective theory of MSLRMs with a
renormalizable is the MSSM with exact R parity so 
the Lightest Supersymmetric Particle (LSP) is stable }}. 

Consider next the detailed structure of MSLRMs. Firstly , just
as for the MSSM, the non zero $B-L$ charges $(+2,-2)$ of the
$\Delta,\Delta^c$ force one into introduce partners
${\bar{\Delta}}, {\overline{\Delta^c}}$ for both with opposite B-L to
cancel gauge anomalies due to their fermionic components. 
Then if one attempts to build a minimal renormalizable model
with just these fields one finds that  one cannot introduce
$\Delta$ interactions in the superpotential with renormalizable
terms only. Thus  LR SSB is impossible. Historically the way around this
was
thought \cite{cvetic}
to be the introduction of a parity odd gauge singlet field. In a
surprising reanalysis \cite{kuchimoha} it was pointed out that
in that model , in the generic case where $M_R>> M_S$ , there
was a circular flat direction passing through the LR breaking
SUSY vacuum and that once soft susy breaking corrections were
switched on the vacuum inevitably preferred to settle in a
charge breaking direction. Thus it was concluded that either one
must give up renormalizability or one must settle for the
restriction to the corner of parameter space where $M_R\sim
M_{B-L}\sim M_W$ , in which case it also followed that R-parity
must be spontaneously broken. The latter alternative is now clearly
quite unacceptable .  In a series of papers \cite{abs,ams,amrs}
the generic case of large $M_R$ was reanalyzed with particular
attention to the structure of the SUSY vacuum using the powerful
theorems available to characterize SSB in SUSY models. This has
established the fields of generic MSLRMs 
(in addition to the supersymmetrized anomaly free set of fields
of the LR model $(Q,Q^c,L,L^c,\phi,\Delta,{\bar{\Delta}},{\Delta}^c
{\overline{\Delta}}^c)$ ) as being one of the following :
 
{a)} Introduce a pair $\Omega(3,1,0,1)\oplus
\Omega^c(1,3,0,1)$ 0f $SU(2)_{L/R}$ triplet fields . Then one can
achieve SSB of the LR symmetry via the $\Omega$s and separately
the SSB of the $B-L$ symmetry, at an independent scale $M_{B-L}$
 by the $\Delta^c,{\bar\Delta}^c$ fields.

{b)} Stay with the minimal set of fields , but (reasoning
that small non-renormalizable corrections must be counted when
the leading effects are degenerate) include the next order $d=4$
operators allowed by gauge invariance in the superpotential.
These operators are of course suppressed by some large scale M
and may be thought to arise either from Planck scale physics
$(M\sim M_{Planck})$ or when one integrates out heavy fields in
some GUT in which the LR model is embedded ($M\sim M_X)$. The 
principal effect of allowing such terms is that the charge
breaking flat direction is lifted and one obtains a
phenomenologically viable low energy effective theory (with
characteristic additional fields at the scale $M_{B-L}^2/M_R$ :
see below). 

{c)} Finally one may introduce a parity odd singlet in
either of cases a) b) which we shall for convenience refer to as
cases $a')$ and $b')$. This case is not quite academic or non-minimal since
such parity odd singlets arise very naturally when one embeds
these models in $SO(10)$.

Although these models contain a very large number of scalar
fields the rigorous study of the structure of their vacuum is
facilitated by a powerful theorem \cite{luty} applicable
to SUSY vacua . Recall that the potential of a supersymmetric
gauge theory has the positive definite form :

\begin{eqnarray}
V_{SUSY} &=& \sum_{Chiral} |F_i|^2 +\sum_{Gauge}
D_{\alpha}^2\nonumber \\
&=& \sum_i |{{\partial W(\phi_i)}\over{\partial \phi_i}}|^2 + 
{1\over 2}\sum_{\alpha} (g_{\alpha}\phi^{\dagger} T_{\alpha} \phi)^2
\label{vsusy}
\end{eqnarray}

Thus the minimization involves finding the set of field values
for which the D and F terms for the different gauge generators
$\alpha$ and complex scalar fields $\phi_i$ vanish. One then has
the following remarkable result :

{\bf{Theorem :}} a) the set of all independent gauge invariant
holomorphic invariants $x(\phi)_a,{\bar{x}}(\phi^{\dagger})^a$
 formed from the chiral fields furnish (complex)
coordinates for the manifold of D-flat vacua .

b) The invariants ${x_a}$ left undetermined by the conditions
$F(\phi)_i=0 $ are coordinates for the space of vacua that are
both F and D flat i.e for the space of SUSY vacua.

A phenomenologically acceptable 
SUSY vacuum must be isolated from other vacua by barriers which
prevent its decay into those vacua. Thus the existence of flat
directions in field space (undetermined holomorphic invariant) 
connecting it with unacceptable vacua is not acceptable. On the
other hand since the vacuum cannot break colour and electric
charge it follows that the soft SUSY breaking terms must
provide positive masses to scalars so that this disaster does
not occur.

With the theorem one can characterize the flat directions out of
the LR symmetry breaking vacuum \cite{amrs} and show that they
violate R-Parity only if they also break charge and hence must
be prevented from doing so by the soft susy breaking terms. Then it
follows that the supersymmetric LR asymmetric vacua are isolated and the
argument given above for the preservation of R-parity at all scales
goes through without any difficulty. The detailed analysis of
symmetry breaking also allows one to calculate the mass spectrum
of the theory. Besides, the usual particles of the SM and their
superpartners at $M_S$ one finds  that certain
superfields associated with $SU(2)_R\times U(1)_{B-L}$ breaking
remain relatively light and for favourable values of the
parameters may even be detectable at current or planned
accelerators. Thus in cases $a)$ and $a')$ one finds that a complete
supermultiplet with the quantum numbers of $\Omega(3,1,0,1)$
has a mass $M_{B-L}^2/M_R$ . If $M_{B-L}<<M_R$ then these
particles could be detectable.  However given the expectation of
$M_{B-L} > 10^{14} GeV$ from neutrino mass this does not appear
to be a likely possibility. In Cases $b)$ and $b')$, which one
may consider as the truly minimal alternative one finds instead
that the entire slew of fields
$\Delta,{\bar{\Delta}},\delta^c_{--},{\bar{\delta} }^c_{++}, 
\delta^c_{0}+ {\bar{\delta} }^c_{0}, H_u', H_d'$ have masses
$\sim M_R^2/M$. If , for instance, $M\sim 10^{19} GeV$ and
$M_R\sim 10^{11} GeV$ then these particles could conceivably be
detectable specially because they include exotic particles
with charge 2 which are coupled to the usual light fermions of
the model.

In the above $H_u',H_d'$ are a pair Higgs doublets left
over after a fine-tuning to keep one pair of doublet superfields
light out of the four (i.e two bidoublets) that must be
introduced to allow sufficient freedom in the Yukawa couplings .
In other words , with a single bidoublet field the restrictive
form of the superpotential ensures that the up and down quark
mass matrices are proportional, which is not acceptable. However
a non trivial feature of these models is that the symmetry
breaking at the Right handed scales furnishes vevs that can
discriminate between $SU(2)_L$ doublets with $T_{3R}=\pm 1/2$.
With two bidoublets one can then ensure that the Yukawa coupling
matrices couplings of one pair of light doublets are not
mutually proportional.

In an interesting pair of papers  following \cite{ams} (which originally 
pointed out the possibility of light doubly
charged particles in the MSLRM) , Mohapatra and collaborators
\cite{mohadatta,mohachacko} analyzed the phenomenology of
light doubly charged particles and also  extended the argument to
include the lepto-quark Higgs in the Pati-Salam GUT . They find
that if doubly charged lepto-quark scalars with masses $\sim 100
GeV$ exist they will mediate exotic scattering processes such as
$\mu^+e^- \longrightarrow \mu^-e^+$ with cross sections in the
picobarn range and hence may be detectable at upcoming
detectors.  In the Pati-Salam
case they find that entire lepto-quark $(3,1,10)$ multiplets can remain
light and thus give rise to measurable rates for Neutron- antineutron
oscillations described by the effective operator
$u_c^2d_c^4$ which arises via the exchange of $\Delta$s .
 This yields a value for the oscillation time 

\begin{equation}
\tau_{N{\bar{N}}} \sim {{\lambda <\Delta^c>
f^3}\over{M_{d^c{\tilde{d}}^c} M_{u^c{\tilde{u}}^c}}}\sim 10^9 -
10^{12} sec
\label{osctime}
\end{equation}

which may be detectable by upcoming experiments at Oakridge
\cite{oakridge}. 

\section{LR SUSY GUTS}

As we have seen, LR SUSY models are natural candidates for SUSY
unification which accommodates neutrino mass. Thus it is natural
to consider further unification in which the various factors of
the LR symmetry group are unified with each other. The two most 
appealing possibilities are unification within the Pati-Salam
group $\patisalam $ and $SO(10)$. The multiplets
{\bf{45,210}} of $SO(10)$ contain parity odd singlets
\cite{mohaparid} and the Pati-Salam gauge group is a subgroup of
$SO(10)$. Thus the study of $SO(10)$ unification 
teaches one much about the Pati-Salam
case as well. Therefore we\cite{abmrs} have re-examined  SO(10) SUSY
unification \cite{dimowil,amso10,babumohap}
 keeping in view the progress in understanding of LR
SUSY models detailed above and developed a minimal
SO(10) Theory of R-Parity and Neutrino mass with the
appealing features of automatic R-parity conservation . A
detailed and explicit study of the SSB at the GUT scale and
various possible intermediate scales was performed. The  mass
spectra in various cases could be  explicitly computed. In
particular the pseudo-goldstone supermultiplets with possibly
low intermediate scale masses ($\sim M_R^2/M_{PS}, M_{PS}^2/M_X$
etc.) that often arise in SUSY GUTS (see \cite{amso10} for an early
example involving SO(10)) were determined. With these computed
(rather than assumed spectra) a preliminary one-loop RG survey
of coupling constant unification in such models was carried out.

In $SO(10)$ matter parity M is a finite gauge symmetry, since under M

\begin{eqnarray}
16 \; \stackrel{M}{\longrightarrow} \;  -16 \quad
10 \; \stackrel{M}{\longrightarrow} \;  10
\label{mparity}
\end{eqnarray}

\noindent and all other representations built out of the fundamental $10$,
such as $45, 54, 126$, etc. are even. The symmetry in (\ref{mparity}) 
is simply $C^2$, where $C$ is the center of $SO(10)$, so that under
it $16 \to i 16$, $10 \to -10$. 
This points strongly towards using a $126$ -dimensional Higgs for the
breaking
of $B-L$ and the see-saw mechanism.

We wish to construct a renormalizable $SO(10)$ theory with a see-saw, and 
this requires the minimum set of Higgs representations which break $SO(10)$
down to the MSSM:

\begin{equation}
S = 54 \; \quad A = 45 \; , \quad \Sigma = 126 \; , \quad \bar 
\Sigma= \overline{126} 
\label{higgses}
\endequation

Although $SO(10)$ is anomaly-free one has to use both 
$\Sigma$  and $\bar \Sigma$ in order to ensure the flatness  of the D-piece
of the potential at large scales $\gg M_W$

The $\patisalam$ decompositions of the $SO(10)$ multiplets we use
are as follows :

\begin{eqnarray}
\psi &= 16 &= Q(2,1,4) + Q^c(1,2,{\bar 4}) \nonumber \\
S & = 54 &= (1,1,1) + (1,1,20) +(3,3,1) + (2,2,6) \nonumber \\
A & = 45  & = \sigma (1,1,15) + \Omega(3,1,1) + \Omega^c (1,3,1) + (2,2,6) 
\nonumber \\ 
\Sigma  & = 126 & = \Delta (3,1,{\overline {10}}) + \Delta^c (1,3,10) +
\phi' (2,2,15) +H_C'(1,1,6) \nonumber \\
{\bar \Sigma} & ={\overline {126}} &= {\bar{\Delta}} (3,1,{ 10})
+ {\overline{\Delta^c}}(1,3,{\overline 10}) + \phi'(2,2,15)
+{\overline H_C}'(1,1,6) \nonumber \\ 
H & = 10 & = \phi (2,2,1) + H_C (1,1,6) \nonumber \\
\end{eqnarray}

The most general superpotential one may build from the fields
$S,A,\Sigma,{\bar{\Sigma}}$ involved in High scale gauge
symmetry breaking is :

\begin{eqnarray}
W &= &{m_S \over 2} \rm{Tr}\, S^2 + {\lambda_S \over 3} \rm{Tr}\, S^3
+ {m_A \over 2} \rm{Tr}\, A^2  + \lambda \rm{Tr}\, A^2S \nonumber \\
& & + m_\Sigma \Sigma \bar\Sigma + \eta_S \Sigma^2 S + \bar\eta_s
{\bar\Sigma}^2 S + \eta_A \Sigma\bar\Sigma A
\label{superpot}
\end{eqnarray}

Assigning vevs to suitable submultiplets as :

\begin{eqnarray}
s &=& \langle (1,1,1)_S \rangle  \nonumber \\
a &=& \langle (1,1,15)_A \rangle \nonumber \\
b &=& \langle (1,3,1)_A \rangle \nonumber \\
\sigma &=& \langle (1,3,10)_\Sigma \rangle \nonumber \\
\bar\sigma &=& \langle (1,3,\overline 10)_{\bar\Sigma}\rangle 
\end{eqnarray}

one can achieve two representative and interesting symmetry breaking chains
which moreover present structural features in counterpoint. They  are :

{\bf (a) :} The case where $SU(2)_R$ is broken before the $SU(4)_c$ :
\begin{eqnarray}
SO(10) & \stackrel{\langle (1,1,1)_S \rangle = M_X }{\longrightarrow} &
G_{PS}= \patisalam \times D_{LR} \nonumber \\
& \stackrel{\langle (1,3,1)_A \rangle = M_R }{\longrightarrow} &
SU(2)_L \times U(1)_R \times SU(4)_C  \nonumber \\
& \stackrel{\langle (1,3,10)_\Sigma \rangle = \langle (1,3, 
{\overline{10}})_{\bar
\Sigma} \rangle = M_{PS} (M_{BL}) }{\longrightarrow} &
\stmd 
\label{chain2}
\end{eqnarray}
In this case the renormalizable LR model without a parity singlet
(case a)) is embedded in the Patisalam model and that in SO(10).

{\bf (b) :} In the other case $SU(4)_c$ is broken simultaneously
with the breaking of the LR discrete symmetry while preserving
the rest of the LR gauge group :
\begin{eqnarray}
SO(10) & \stackrel{\langle (1,1,1)_S \rangle = M_X }{\longrightarrow} &
G_{PS} =\patisalam \times D_{LR} \nonumber \\
& \stackrel{\langle (1,1,15)_A \rangle = M_{PS} }{\longrightarrow} &
\leftright \nonumber \\
& \stackrel{\langle (1,3,10)_\Sigma \rangle = \langle (1,3, 
{\overline{10}})_{\bar\Sigma} \rangle = M_R }{\longrightarrow} &
\stmd 
\label{chain1}
\end{eqnarray}

In SO(10) GUTS the discrete LR symmetry may be naturally embedded in the
gauge group . In fact $D_{LR} = \sigma_{23} \sigma_{67}$ . Then
it follows that the  $\sigma(1,1,15)$
submultiplet of the {\bf{45}} multiplet contains a SM  singlet
which is parity odd. Thus when this singlet is used for the
second stage of SSB, LR symmetry is broken even though the gauge
group is still $\leftright$. So this case corresponds to
embedding SUSY LR with a POS into $\patisalam$ and that into
SO(10). Integrating out the fields left heavy after GUT scale
symmetry breaking down to the PS gauge group gives a non-
renormalizable superpotential involving only the $(1,1,15)_A,
(1,3,10+{\overline{10}}), (3,1,10+{\overline{10}})$ which is the PS
generalization of the non renormalizable model with a parity odd
singlet (Case b')) .

Notice that one needs both the multiplets A(45) and the multiplet
$54$. For if one drops the symmetric multiplet (54) one finds
that the vevs necessarily preserve SU(5). While if one drops 
the antisymmetric multiplet A(45) one can break SO(10) to $G_{PS}$
but the vev of the 126 and ${\overline{126}}$ multiplets vanishes .

The non-trivial F equations are 
\begin{eqnarray}
F_{(1,1,1)_S} = m_s s - {1 \over 2} \lambda_S s^2 + {2\over 5}
 \lambda
 (a^2 -b^2) &=&0 \nonumber \\
F_{(1,1,15)_A} = m_A a + 2 \lambda_A a s + {1 \over 2} \eta_A
\sigma\bar\sigma
&=& 0 \nonumber \\
F_{(1,3,1)_A} = m_A b - 3 \lambda_A b s + {1\over 2} \eta_A
\sigma \bar\sigma 
&=& 0 \nonumber \\
F_{(1,3,10)_\Sigma} = \sigma \left[ m_\Sigma + \eta_A (3 a + 2 b) \right]
&=& 0 \nonumber \\
 F_{(1,3,10)_{\bar\Sigma}} =\bar \sigma \left[ m_\Sigma + 
\eta_A (3 a + 2 b) \right]
&=& 0
\label{efeqs}
\end{eqnarray}

While the D equations demand only $\sigma =\bar \sigma$. 

The choice of the ratio $a/b$ determines the chain of breaking.
Note that $a,b$ must {\it{both}} be nonzero for the 
$\sigma{\bar\sigma}$ to be non zero. Thus there is no Dimopoulos Wilczek
mechanism here unless $M_R^2/M_X \sim M_W $ or $M_{PS}^2/M_X\sim M_W$ .

\begin{description}
\item[(a)] $s\sim M_X \gg b \sim M_{R}\gg \sigma= \bar\sigma \sim M_{PS} 
\gg a \simeq M_{PS}^2/M_X$ corresponding to the chain of
eqn(\ref{chain2})  is achieved by fine-tuning
\end{description}
\begin{equation}
m_A - 3 \lambda_A s \simeq {\sigma^2 \over b} \ll s
\label{finea}
\end{equation}
which then ensures 

\begin{equation}
a \simeq {\sigma^2 \over s} \ll \sigma
\label{finea2}
\end{equation}

$s\sim M_X \gg b \sim M_{R}\gg \sigma= \bar\sigma \sim M_{PS}=M_{B-L} 
\gg a \simeq M_{PS}^2/M_X$.

\begin{description}
\item[(b)] Similarly for  Case(b) one interchanges the roles of $a$ and
$b$.

$s\sim M_X \gg a \sim M_{PS}\gg \sigma= \bar\sigma \sim M_R 
\gg b \simeq M_R^2/M_X$ corresponding to the chain of
eqn(\ref{chain1})  is achieved by fine-tuning
\end{description}
\begin{equation}
m_A + 2 \lambda s \simeq {\sigma^2 \over a} \ll s
\label{fineb}
\end{equation}
which then ensures 

\begin{equation}
b \simeq {\sigma^2 \over s} \ll \sigma
\label{fineb2}
\end{equation}

 Note that we have ignored the vevs of the 
superpartners of the matter fields in the  $ 16$  along with the vevs 
$<10_H>$ since these would break the SM symmetries which we know
to be preserved to far below the scales under discussion.
Moreover the couplings $W=\psi \psi H + \psi \psi {\bar\Sigma}...$
 in the superpotential imply that if $10$ vev 
is negligible then the $16$ vev (in the ${\tilde \nu}^c$ direction; the
other vevs break SM symmetries) is zero if $\overline 126$ vev is not.
So that the protection of $<{\tilde\nu}_c> =0$ carries over from
the LR symmetric case.

With these solutions of the SUSY potential minimization
equations in hand we can calculate the mass spectrum of the
theory . One obtains \cite{abmrs} in case a)

\begin{center}
\framebox{ \begin{tabular}{l|l}
\hspace{2cm}State & \hspace{1cm} Mass \\
  \hline & \\ 
 \begin{tabular}{l}
all of $S$ in 54 \\
all of $A$ in 45, except $(3,1,1)_A + (1,3,1)_A$\\
all of $\Sigma$ in 126 + $\bar \Sigma$ in $\overline{126}$,
 except  $SU(4)_C$ decuplets
\end{tabular}
&  $\sim M_X$ \\
 & \\ \hline & \\
 \begin{tabular}{l}
$(3,1,\overline{10})_\Sigma$  + $(3,1,{10})_{\bar\Sigma}$  \\
$(1,\stackrel{0}{-},\overline{10})_\Sigma$  and
 $(1,\stackrel{+}{0},{10})_{\bar\Sigma}$  \\
$\omega_c^\pm$ from $(1,3,1)_A$
\end{tabular}
&  $\sim M_{R}$ \\
 & \\ \hline & \\
 \begin{tabular}{l}
color triplets and singlets from  \\
\hspace{0.5cm}$(1,+,\overline{10})_\Sigma$  and $(1,-,{10})_{\bar\Sigma}$
\end{tabular}
&  $\sim M_{PS}$ \\
 & \\ \hline & \\
 $(3,1,1)_A$ &
 $\sim Max\left[{ M_R^2\over M_X}, {M_{PS}^2\over M_R} \right]$\\
 & \\ \hline & \\
 \begin{tabular}{l}
color sextets from \\
 \hspace{0.5cm}$(1,+,\overline{10})_\Sigma$  and $(1,-,{10})_{\bar\Sigma}$
\end{tabular}
&  $\sim {M_{PS}^2 / M_X}$\\
&  
\end{tabular}}

\vspace{0.5cm}
{\bf Table 1}: Mass spectrum for the symmetry breaking chain
$SO(10) \stackrel{M_X}{\rightarrow}\patisalam \stackrel{M_{R}}
{\rightarrow}
SU(2)_L\times U(1)_R\times SU(4)_C  \stackrel{M_{PS}}{\rightarrow}
 \stmd $.
The states in $(1,3,10) $ and $(1,3,\overline{10})$ were decomposed 
according to their $T_{3R}$ number, for example $(1,+,10)$ denotes 
the component of $(1,3,10)$ with $T_{3R} = +1$, etc. 

\end{center}
\vfil
\eject
Similarly  in case b) we get 
\begin{center}
 \framebox{\begin{tabular}{l|l}
\hspace{2cm}State & \hspace{1cm} Mass \\
\hline 
 & \\
 \begin{tabular}{l}
all of $S$ in 54 \\
all of $A$ in 45, except $(1,1,15)_A$\\
all of $\Sigma$ in 126 + $\bar \Sigma$ in $\overline{126}$,
 except  $SU(4)_C$ decuplets\\
Color triplets $H_C(1,1,6)$ in 10.
\end{tabular}
&  $\sim M_X \sim<S> $\\
 & \\
\hline
 & \\
 \begin{tabular}{l}
$(3,1,\overline{10})_\Sigma$  + $(3,1,{10})_{\bar\Sigma}$  By $D_{LR}$
breaking\\
color triplets and sextets of$(1,3,\overline{10})_\Sigma$  and
 $(1,3,{10})_{\bar\Sigma}$  \\
color triplets of $(1,1,15)_A$ (Higgs)
\end{tabular}
&  $\sim M_{PS} \sim<A>$\\
% & \\ \hline & \\
 \begin{tabular}{l}
$\nu^c  $\\
$\delta_c^0 + {\bar \delta}_c^0, \delta_c^\pm,\bar\delta_c^\pm  $\\
\hspace{0.5cm} from the color singlets of $(1,3,\overline{10})_\Sigma$
  and $(1,3,{10})_{\bar\Sigma}$ 
\end{tabular}
&  $\sim M_{R} \sim <\Sigma>$\\
%& \\ \hline & \\

Color octet and singlet in $(1,1,15)$ 
&  $\sim Max\left[ {M_R^2\over M_{PS}}, {M_{PS}^2\over M_X} \right]$ \\
%& \\ \hline & \\
 \begin{tabular}{l}

$\delta_c^{--},{\bar \delta}_c^{++},\delta_c^0 - {\bar \delta}_c^0 $ 
\\
\hspace{0.5cm} from the color singlets of $(1,3,\overline{10})_\Sigma$ 
 and $(1,3,{10})_{\bar\Sigma}$ 
\end{tabular}
&  $\sim {M_{R}^2 / M_X}$ \\
 & \\
\end{tabular}}

\vspace{0.5cm}
{\bf Table 2}: Mass spectrum for the symmetry breaking chain
$SO(10) \stackrel{M_X}{\rightarrow}\patisalam
\stackrel{M_{PS}}{\rightarrow}
\leftright  \stackrel{M_R}{\rightarrow} \stmd $
\end{center}

Notice particularly how the left handed $\Delta$s become superheavy 
($M\sim M_X$) while the right handed ones do not become superheavy 
due to breaking of  
$D_{LR}$. Given these mass spectra one may carry out the RG
analysis of the gauge coupling evolution . As data one has the
values $\alpha_i(M_Z) ,i=1,2,3$ while the mass scales
$M_S,M_{PS},M_R,M_X$  in Case a) and $M_S,M_{B-L},M_{PS},M_X$ in
Case b) together with the value of the coupling constant at
unification $\alpha_U$ are unknowns.
Since the deviation of $log M_S$ from $log M_Z$ cannot
be very large it may be ignored in an approximate one-loop
analysis. Then one obtains a one parameter family of solutions
for the mass scales of interest. For instance in Case b)
one obtains for the case of two light Higgs doublets ($t=log_{10}(M/GeV)$)

\begin{equation}
t_X=16.4 t_{PS}=14.7 +{.07/{\alpha_U}},t_R=13.9 +{.1\over \alpha_U}
\label{RGres1}
\end{equation}

while consistency of the assumptions made regarding the relative
magnitudes of the intermediate geometric scales
($M_{PS}^2/M_X,M_R^2/M_{PS}$) requires that $\alpha_U^{-1} < 23.6$

If one assumes that an additional pair of Higgs doublets
begins to contribute above the scale $M_R^2/M_X$
 (as indicated by the symmetry breaking in
case b) of the LR SUSY model) then one finds (here ${\bar t}=
log_{10} Q$) :

\begin{equation}
t_X=14.2 +{.09/{\alpha_U}};\quad t_{PS}=12.7
+{.15/{\alpha_U}},\quad t_R=9.9 +{.28\over \alpha_U} 
\label{RGres2}
\end{equation}
Here consistency requires that $\alpha_U^{-1} < 22.7$ while the
current bounds on proton decay mediated by gauge particles
imply that $t_X > 15.5$ so that $t_P>14.7 , t_R > 13.5$.
Similar results obtain in Case a) . Generically requiring $t_X
\geq 15.5$ while $M_R,M_{PS},M_{B-L} \geq 10^{13} GeV$ or so.
Even given the uncertainties of the analysis in which we have
ignored threshold and 2-loop effects etc. the qualitative
conclusion is clear:  these models are victims of a sort of
``SU(5) conspiracy'' as far as the gauge breaking chain is
concerned since the intermediate scales are so close to the GUT scale.

\section{Fermion Masses and Proton Decay}
The complicated question of realistic mass spectra in the class of SO(10)
SUSY GUTS we have focused on  has
been considered by several sets of authors following the work of Babu
and Mohapatra \cite{babumohap,odatak}. As
usual in GUTS the fermion masses derived from the GUT dictated
relations between Yukawas must be RG improved by running
down to EW scales.
Making the assumption that the light bidoublets that give masses
to the charged fermions are in fact a mixture of the bidoublets
contained in the 10 plet and ${\overline{126}}$ Higgs which have
suitable Yukawa couplings it has been found that even without
the freedom allowed by the Type II seesaw mechanism for Neutrino
mass it is possible to fit the observed fermion masses if one takes the
light bidoublets to be a mixture of those contained in two 10 plets
and a single ${\overline126}$ plet provided the LSND data are
discounted. However the precise way in which this light mixture
arises (while also making all colored fields superheavy) has not
been worked out. 

Since the generation-wise freedom to choose the
values of the Yukawa couplings continues to be present, SO(10)
alone does not shed much light on the pattern of Fermion masses.
However recent work \cite{bpw} has pointed out interesting
connections between seesaw neutrino masses and proton decay via
$d=5$ operators in SUSY SO(10) GUTS. Recall that in traditional
GUTS the exchange of superheavy gauge bosons mediates proton
decay leading to a 4 fermion $\Delta B= \Delta L =1$ operator
($qqql$) with coefficient $\sim g^2/M_X^2$ . This implies a
nucleon decay dominantly in the $\pi e$ channel with a lifetime
$\sim 10^{28\pm 1} (M_X/10^{14.5} GeV)^4$ yrs. Since the current
limits on Nucleon decay lifetime $\tau_N$ 
via these channels are $> 10^{33} yrs$
 \cite{superk} it follows that the non-supersymmetric case is
contraindicated. On the other hand since $M_X\sim 10^{16}$ in
the minimal SUSY SU(5) GUT , it is compatible with this limit,
as is any theory which reduces to it at a sufficiently high
scale (like the examples above). It is perhaps worth remarking
that with the lower values of $M_X$ possible in the SO(10) case
the $d=6$ nucleon decay operators with their characteristic
flavour diagonal decay modes may be observable in contrast to
the minimal SUSY SU(5) case.

However in SUSY theories there is a much faster source of
nucleon decay via $d=5$
operators\cite{weind5,dimoell,hisano} 
that arise from the exchange of the color triplet partners of
the EW Higgsinos between two fermions and two sfermions. After dressing by
gaugino exchange to convert the scalars into fermions one
obtains the effective four fermi operator but with a coefficient
which is uncertain due to the uncertainty in SUSY breaking
parameters and in $tan\beta $.For low to moderate $tan\beta$
wino dressed diagrams are dominant over those with dressing by
other gauginos. Since constraints of Bose symmetry and color
antisymmetry imply that the 4 fermi operators must be flavor 
non-diagonal the dominant decay of nucleons is 
($N\rightarrow K \nu_{\mu}$ ) with  life
times \cite{hisano} :
\begin{eqnarray}
\tau(p &\rightarrow& K^+ {\bar{\nu}}_{\mu}) \sim |{{M_{H_C}}\over
{10^{17} GeV}} {{M_S\over { GeV}}} |^2 10^{32} yrs \nonumber \\
\tau(p & \rightarrow & K^+ {\bar{\nu}}_{e}) \sim |{{M_{H_C}}\over
{10^{17} GeV}} {{M_S\over { GeV}}} |^2 10^{33} yrs \nonumber \\
\tau(n &\rightarrow & K^0 {\bar{\nu}}_{\mu}) \sim |{{M_{H_C}}\over
{10^{17} GeV}} {{M_S\over { GeV}}} |^2 10^{31.5} yrs \nonumber \\
\end{eqnarray}

Charged lepton modes are suppressed relative to these neutrino 
modes by $10^4$ or more. Current experimental limits are in
the region of $10^{33} yrs$ and improving. Thus $d=5$
decay modes represent a strong constraint on SUSY GUT models
which are already (at least for minimal models ) on the verge of
being ruled out.  For large $tan \beta $ \cite{babubarr} gluino
and Higgsino dressed diagrams which are flavor suppressed 
can become important leading to much larger charged lepton decay
rates making them comparable with the dominant modes in the low
$tan \beta $ case. Thus, if observed , nucleon decay can provide
insight into the EW symmetry breaking in SUSY theories and in
particular into the value of $tan\beta $.

Finally in a notable pair of papers Babu , Pati and Wilczek
\cite{bpw} have recently pointed out that in SO(10) theories with either a
renormalizable (mediated by $W= {\overline{126}}_H 16_M 16 _M +...$
 or a non renormalizable
(mediated by $W=({\overline{16}}_H)^2 16_M^2/M + ...$) seesaw mechanism for
neutrino mass there exist $d=5$ nucleon decay operators that
arise via exchange of color triplet Higgsinos from the $\overline
126$ or $\overline 16$ respectively.
 The strength of these operators is thus directly linked 
to the Yukawa coupling ($f\sim m_{Dirac} ^2/(m_{\nu_L} <\Delta^c>)$ )
between the seesaw Higgs and the matter fields. Assuming
the neutrino Dirac masses at the GUT scale take the values
suggested by SO(10) ($\sim 1-10 MeV, 1-4 GeV, 100-120 GeV$)
and a suitable right handed Majorana neutrino mass matrix 
to ensure compatibility of the see-saw masses with the
values suggested by the Atmospheric and solar neutrino data
gives $f_{33}\sim 5 \times 10^{-2} $ . A detailed analysis then shows
that even at low $tan\beta$ the charged lepton $d=5$ decay modes
can become comparable to the neutrino final state modes.
Moreover the lifetimes suggested are on the borderline of
conflict with the current experimental limits. Thus this work
has opened an interesting and amazing direct connection between
the seesaw mechanism for neutrino mass and nucleon decay in GUTS
: two topics that earlier were thought to be quite separate. 

It is worth noting here that the classic mechanisms for
cosmic baryogenisis via GUT scale baryon number violation have
fallen into disfavor since it was realized that the standard
model leads to unsuppressed $B+L$ (but not $B-L$) violation at
high temperatures due to non perturbative processes . However
these very processes can be enlisted\cite{leptogenisis}
 to provide a very robust
mechanism for Leptogenisis at high scales based on the lepton
number and CP violating out of equilibrium decay of 
right handed neutrinos at temperatures $> 10^{10} GeV$. This
lepton asymmetry is then equilibrated by non-perturbative
processes into a Baryon and Lepton asymmetry.

To summarize 

$\bullet $ There is a clear logical chain leading from the SM with
neutrino mass to the Minimal Supersymmetric LR models with
renormalizable seesaw mechanisms developed in detail recently.

$\bullet $ These MSLRMs have the MSSM with R parity and  seesaw
neutrino masses and have quasi exact B-L as their effective low
energy theory. They can also have light Higgs
triplet supermultiplets in their low energy spectra leading to
very distinctive experimental signatures.

$\bullet $ They can be embedded in GUTS based on the PS group or
SO(10). The former case may be more suitable in stringy
scenarios which so far disfavor light (on stringy scales) SO(10)
GUT Higgs of dimension $>54$. 

$\bullet $ The SSB in the SO(10) SUSY GUT has been worked out
explicitly and the mass spectra calculated. This allowed us to
perform a RG analysis based on calculated spectra leading to the
conclusion that $M_X\geq 10^{15.5} GeV$ while $M_R,M_{PS}\geq
10^{13} GeV$. With $M_X$ at its lower limit the $d=6$ operators
for nucleon decay can become competitive with the $d=5$
operators raisining the possibility that observation of
$p-->\pi^{0}e^+$ need not rule out SUSY GUTS after all.

$\bullet $ Fermion mass spectra can be compatible with charged
fermion mass data and neutrino mass values suggested by neutrino
oscillation data from super Kamiokande and Solar neutrino
oscillation experiments .

$\bullet $ Dimension five operators in theories with seesaw lead to
a remarkable connection between neutrino masses and nucleon
decay which constrains these models fairly tightly and makes
them testable by upcoming nucleon stability measurements.

$\bullet $ Further work on the doublet triplet splitting problem,
the question of fermion mass spectra , two loop RG analysis etc. 
is required.

$\bullet $ {\it{ Thus LR SUSY seesaw models and their GUT
generalizations look good. The show has just begun but it aint
over till the f... neutrino sings ........!}}

\endgroup
\end{document}